\documentstyle[12pt]{article}

\newcommand{\beq}{\begin{equation}}
\newcommand{\eeq}{\end{equation}}
\newcommand{\beqa}{\begin{eqnarray}}
\newcommand{\eeqa}{\end{eqnarray}}
\newcommand{\beqar}{\begin{eqnarray*}}
\newcommand{\eeqar}{\end{eqnarray*}}

\newcommand{\Ga}{\Gamma}

\newcommand{\inn}{\!\cdot\!}

\renewcommand{\l}{\lambda}

\newcommand{\z}{\zeta}

\newcommand{\eg}{{\it e.g.,}\ }
\newcommand{\ie}{{\it i.e.,}\ }
\newcommand{\labell}[1]{\label{#1}} %{\label{#1}} %
\newcommand{\reef}[1]{(\ref{#1})}
\newcommand\prt{\partial}

\newcommand\cL{{\cal L}}

\newcommand\Tr{{\rm Tr}}
\begin{document}

\thispagestyle{empty} \rightline{\small  \hfill IPM/P-2003/035}
\vspace*{1cm}

\begin{center}
{\bf \Large
S-matrix elements and off-shell tachyon action \\
with non-abelian gauge symmetry  \\

 }
\vspace*{1cm}

{Mohammad R. Garousi}\\
\vspace*{0.2cm}
{ Department of Physics, Ferdowsi university, P.O. Box 1436, Mashhad, Iran}\\
\vspace*{0.1cm}
and\\
{ Institute for Studies in Theoretical Physics and Mathematics
IPM} \\
{P.O. Box 19395-5531, Tehran, Iran}\\
\vspace*{0.4cm}

\vspace{2cm}
ABSTRACT
\end{center}
We propose that there is a unique expansion for the string theory
S-matrix elements of tachyons that corresponds to non-abelian
tachyon action. For those S-matrix elements which, in their
expansion, there are  the Feynman amplitudes resulting from the
non-abelian kinetic term,  we give a prescription on how to find
the expansion. The gauge invariant action is an $\alpha'$
expanded action, and the tachyon mass $m$ which appears as
coefficient of many different couplings,
is arbitrary.\\
We then analyze in details the S-matrix element of four tachyons
and the S-matrix element of two tachyons and two gauge fields, in
both bosonic and superstring theories, in favor of this
proposal.  In the superstring theory, the leading terms of the
non-abelian gauge invariant couplings are in  agreement with the
symmetrised trace of the direct non-abelian generalization of the
tachyonic Born-Infeld action in which the tachyon potential is
consistent with $V(T)=e^{\pi\alpha' m^2T^2}$. In the bosonic
theory, on the other hand,  the leading terms are those appear in
superstring case as well as some other gauge invariant couplings
which  spoils the symmetrised trace prescription. These latter
terms are zero in the abelian case.
%This study also indicates that the part of the tachyon action
%which is independent of tachyon mass is identical for both tachyon
%and   massless transvers scalars.
%The non-leading terms are
%expected to be related to the terms with higher covariant
%derivative of tachyon or gauge field strength.
%We also analyze the S-matrix element of one closed string
%graviton, one gauge field and one tachyon in bosonic string
%theory. In this case, the amplitude  contains tachyonic pole and
%contact terms   in the non-commutative limit. The tachyonic pole
%and the leading contact term are reproduced by non-commutative
%tachyonic DBI action which includes Wilson line. This indicates
%that the tachyon potential in the tachyonic DBI action has linear
%term. NOT TRUE. THE COUPLING OF OPEN STING TO CLOSED STRING
%IN THE BOSONIC CASE IS NOT CONSISTENT WITH THE WAY CLOSED STRINGS
%APPEARS IN THE DDI ACTION. BUT FOR SUPERSTRING IT IS THE CASE.

\vfill
\setcounter{page}{0}
\setcounter{footnote}{0}
\newpage

\section{The idea} \label{intro}

Decay of unstable branes is an interesting process which might
shed new light in understanding properties of string theory in
time-dependent backgrounds \cite{mgas}-\cite{sen2}. In particular,
by studying the unstable branes in the boundary conformal field
theory (BCFT), Sen has shown that the end of tachyon rolling  in
this theory is a tachyon matter with zero pressure and non-zero
energy density \cite{asen2}. These results is then  reproduced in
field theory by the tachyonic DBI action \cite{mg,ebmr,dk}. The
form of tachyon potential in this action at minimum of potential
is then fixed by using the fact that the higher derivative
corrections to the tachyonic DBI action  are not important at
this point, and the action should not have plane wave solution
\cite{asen3}. Possible application of the tachyonic DBI action to
cosmology have been discussed in \cite{cosmo}.

The tachyonic DBI action was originally proposed as an action
that when transformed under the Seiberg-Witten map, \ie a
non-commutative gauge invariant action, is consistent with the
non-commutative expansion of the S-matrix element of one graviton
and two tachyons in the presence of background B-flux \cite{mg},
\ie an expansion around $s=0$ which produces massless pole as
first leading term and contact term with the $*'$-product as next
leading term\footnote{ The $*'$-product reflects the fact that
the non-commutative gauge invariant coupling of graviton to
D-brane has open Wilson line \cite{mine}-\cite{mg11}.}. The
massless pole is reproduced by non-commutative gauge invariant
kinetic term, and the contact term is reproduced by the
non-commutative tachyon-tachyon-graviton coupling. Both exist in
the above non-commutative action\footnote{Throughout the paper,
we are appealing to the specific meaning of  the tachyon action as
a generating functional for producing the leading terms of the
string theory S-matrix elements.}. The Mandelstam variable $s$ in
the amplitude is arbitrary, however, expansion around any other
point  is not consistent with the non-commutative gauge symmetry,
\eg it does not produce the massless pole, nor the contact term
with $*'$-product. We speculate that this is a general fact for
all other S-matrix elements involving tachyons. That is, there is
only one limit for the S-matrix elements that is consistent with
non-commutative gauge symmetry. We call this limit the
non-commutative limit. Expansion of S-matrix elements at this
limit would produce the Feynman amplitudes that result from the
non-commutative gauge invariant tachyon action.

When there are  $N$ coincident  D-branes the $U(1)$ gauge
symmetry of an individual D-brane is enhanced to a non-abelian
$U(N)$ symmetry  \cite{ew1}. The above proposal in this case is
that there is only one limit for the S-matrix elements involving
tachyons that is consistent with the non-abelian gauge symmetry.
We call this limit the non-abelian limit. The limit would be such
that when the S-matrix elements expanded around it, they
reproduce the Feynman amplitudes resulting from non-abelian gauge
invariant tachyon action. In general, however, it is nontrivial
to find such a limit for a S-matrix element.

For a class of S-matrix elements this limit can be found easily.
This class includes those S-matrix elements that produce, at the
non-abelian limit, massless and/or tachyonic poles which are
related to the  non-abelian kinetic term in field theory side,
\eg the above example \cite{mg} in the non-commutative case.
Since both tachyon and massless transverse scalars transform in
the adjoint representation of $U(N)$ group, they have identical
non-abelian kinetic terms. The Feynman amplitudes resulting from
kinetic terms are, then, similar in both cases. Therefore, the
expansion of  S-matrix element of tachyons and  the expansion of
S-matrix element of scalars should be similar to produce these
Feynman amplitudes. On the other hand, we know the non-abelian
limit of the S-matrix element of the scalars, \ie sending all
Mandelstam variables to zero. Using similar steps for the
S-matrix element of tachyons, one would find the
non-abelian/non-commutative limit of the S-matrix element of
tachyons. The non-commutative limit of the S-matrix element of
four tachyon vertex operators has been found in \cite{mg3} in
this way.

An interesting observation, made in \cite{mg1,mg2}, in the
non-commutative case, is that the S-matrix element of tachyon
vertex operators and the S-matrix element of the massless scalar
vertex operators can be written in a universal/off-shell form. The
non-commutative limit of the S-matrix elements in the universal
form is the limit that all the Mandelstam variables in the
amplitude approach zero. This observation can be made in the
non-abelian case as well.  We interpret this, here, as an
indication that the on-shell mass of tachyon or the scalar does
not appear in the S-matrix element in the universal from.
Accordingly,  the tachyon mass in the non-abelian action  must be
arbitrary/off-shell. The non-abelian tachyon action is an action
expanded in terms of $\alpha'$ which includes  covariant
derivative of tachyon and the off-shell mass. The off-shell mass
may appears as coefficient of many different terms in the action.
If one restricts the arbitrary mass to the on-shell value of
tachyon, \ie $m^2\sim -1/{\alpha'}$, then, the expansion is not an
$\alpha'$ expansion any more. Clearly, for the on-shell massless
scalar case, the action would be the non-abelian
low-energy-effective action of the D-branes.

In this paper, we would like to analyze in details the S-matrix
element of four tachyons,  and the S-matrix element of two
tachyon and two gauge field vertex operators in favor of the
above proposal. We perform this analyzes  in both superstring and
bosonic string theory.  In the superstring case that we perform
the calculations in the next section, we show that the S-matrix
element of four tachyons in the non-abelian limit has massless
poles, and infinite tower of contact terms. And the S-matrix
elements of two tachyons and two gauge fields  has both tachyonic
and massless poles, and infinite tower of contact terms.  The
tachyonic and massless poles are reproduced in field theory side
by the Feynman amplitudes resulting from non-abelian kinetic
term. The leading contact terms, on the other hand, are
reproduced by  some other non-abelian gauge invariant couplings.
We will show that these couplings are fully reproduced by
expansion of the symmetrised trace of the non-abelian tachyonic
BI action with tachyon potential $V(T)=e^{\pi\alpha'm^2T^2}$. In
section 3,  we perform the calculations for the bosonic case. The
amplitudes in this case has two kind of tachyonic poles. In the
non-abelian limit, one of them must be expanded and the other
one  is reproduced by  the standard non-abelian kinetic term. In
this case, the leading contact terms are consistent with the above
non-abelian tachyonic BI action, and  some other gauge invariant
couplings which are zero in the abelian case. Section 4 is
devoted for some discussions on the results.

\section{Amplitudes in superstring theory}

Using the world-sheet conformal field theory technique \cite{jp},
one can evaluate any 4-point functions by evaluating the
correlation function of their corresponding vertex operators.
Performing the correlators, one finds, in general,  that the
integrand has $SL(2,R)$ symmetry. One should fix, then,  this
symmetry by fixing position of three vertices in the real line.
Different fixing of these positions give different ordering of
the four vertices in the boundary of the world-sheet. One should
add all non-cyclic permutation of the vertices to get the correct
scattering amplitude. So one should add the amplitudes resulting
from the fixing $(x_1=0, x_2, x_3=1,x_4=\infty)$, $(x_1=0,
x_2,x_4=1,x_3=\infty)$, $(x_1=0, x_3,x_4=1,x_2=\infty)$,
$(x_1=0,x_3,x_2=1,x_4=\infty)$, $(x_1=0, x_4, x_2=1,x_3=\infty)$,
$(x_1=0,x_4,x_3=1,x_2=\infty)$. After these gauge fixing, one
ends up with only one integral  which gives the beta function. The
S-matrix elements that we will analyze in this paper can be found
in the standard books \cite{mgjs, jp}. The S-matrix elements in
the discussion section which  are not in \cite{mgjs,jp}, can
easily be performed using the above prescription.

\subsection{Four tachyons amplitude}

The S-matrix element of four open string tachyon vertex operators
in the supersting theory   is given by \cite{mgjs,jp} \beqa A^{\rm
tachyon}&\sim&\alpha\frac{\Ga(-2t)\Ga(-2s)}{\Ga(-1-2t-2s)}+
\beta\frac{\Ga(-2s)\Ga(-2u)}{\Ga(-1-2s-2u)}+
\gamma\frac{\Ga(-2t)\Ga(-2u)}{\Ga(-1-2t-2u)}\,\,,\labell{a1} \eeqa
where the Mandelstam variables are \beqa
s&=&-\alpha'(k_1+k_2)^2/2\,\,,\nonumber\\
t&=&-\alpha'(k_2+k_3)^2/2\,\,,\nonumber\\
u&=&-\alpha'(k_1+k_3)^2/2\,\,.\labell{mandel} \eeqa  The
coefficients $\alpha,\beta,\gamma$ are the non-abelian group
factors \beqa \alpha&=&\frac{1}{2}\left(\frac{}{}{\rm
Tr}(\l_1\l_2\l_3\l_4)+{\rm
Tr}(\l_1\l_4\l_3\l_2)\right)\,\,,\nonumber\\
\beta&=&\frac{1}{2}\left(\frac{}{}{\rm Tr}(\l_1\l_3\l_4\l_2)+{\rm
Tr}(\l_1\l_2\l_4\l_3)\right)\,\,,\nonumber\\
\gamma&=&\frac{1}{2}\left(\frac{}{}{\rm Tr}(\l_1\l_4\l_2\l_3)+{\rm
Tr}(\l_1\l_3\l_2\l_4)\right)\,\,.\labell{phase}\eeqa The on-shell
condition for the tachyons are $k_i^2=1/(2\alpha')$, and  the
Mandelstam variables satisfy the constraint \beqa
s+t+u&=&-1\labell{con1}\,\,. \eeqa

Standard non-abelian kinetic term in field theory produces
massless poles in $s$-, $t$-,  $u$-channels. However, the
constraint \reef{con1} does not allow us to sent all $s,t,u$ to
zero at the same time, \ie $s,t,u\rightarrow 0$, to produce
massless poles. In the non-commutative case, it is shown in
\cite{mg3} that in order to produce the massless poles and
contact terms resulting from non-commutative kinetic term, one
should arrange the amplitude in a specific forms, \ie
$A=A_s+A_t+A_u$, and then send $s\rightarrow 0$, $t,u\rightarrow
-1/2$ in $A_s$ to produce massless pole in $s$-channel,
$t\rightarrow 0$, $s,u\rightarrow -1/2$ in $A_t$ to produce
massless pole in $t$-channel, and send $u\rightarrow 0$,
$s,t\rightarrow -1/2$ in $A_u$ to produce massless pole in the
$u$-channel. Similar thing happens here. That is, one should
write $A^{\rm tachyon}=A_s^{\rm tachyon}+A_t^{\rm
tachyon}+A_u^{\rm tachyon}$ where \beqa A^{\rm tachyon}
_s&\sim&\alpha\frac{\Ga(-2s) \Ga(-2t)}{\Ga(-1-2s-2t)}+
\beta\frac{\Ga(-2s)\Ga(-2u)}{\Ga(-1-2s-2u)}\nonumber\\
&&-
\gamma\frac{\Ga(-2t)\Ga(-2u)}{\Ga(-1-2t-2u)}\,\,,\nonumber\\
A^{\rm tachyon} _u&\sim&-\alpha\frac{\Ga(-2s)
\Ga(-2t)}{\Ga(-1-2s-2t)}+
\beta\frac{\Ga(-2u)\Ga(-2s)}{\Ga(-1-2u-2s)}\nonumber\\
&&+
\gamma\frac{\Ga(-2u)\Ga(-2t)}{\Ga(-1-2u-2t)}\,\,,\nonumber\\
A^{\rm tachyon} _t&\sim&\alpha\frac{\Ga(-2t)
\Ga(-2s)}{\Ga(-1-2t-2s)}-
\beta\frac{\Ga(-2s)\Ga(-2u)}{\Ga(-1-2s-2u)}\nonumber\\
&&+\gamma
\frac{\Ga(-2t)\Ga(-2u)}{\Ga(-1-2t-2u)}\,\,,\labell{a717}\eeqa The
field theory massless poles and contact terms resulting from
non-abelian kinetic term are reproduced by sending \beqa s-{\rm
channel}:&&\lim_{s\rightarrow 0\,,t,u\rightarrow -1/2}A_s^{\rm
tachyon}\nonumber\\
t-{\rm channel}:&&\lim_{t\rightarrow 0\,,s,u\rightarrow
-1/2}A_t^{\rm
tachyon}\nonumber\\
u-{\rm channel}:&&\lim_{u\rightarrow 0\,,s,t\rightarrow
-1/2}A_s^{\rm tachyon}\labell{lim1}\eeqa It may seems strange
that in this limit one should send $s$, say, once to zero and once
to $-1/2$. However,  this happens only in the particular form of
the amplitude \reef{a717}. We will show now that one can use the
constraint \reef{con1} to rewrite the amplitude \reef{a717} in
such a way that the limit corresponds to sending all $s,t,u$ to
zero. Consider the gamma functions in the first term in $A_t^{\rm
tachyon}$. In the above limit, it has the following expansion:
\beqa \lim_{t\rightarrow0; s,u\rightarrow
-1/2}\frac{\Ga(-2t)\Ga(-2s)}{\Ga(-1-2s-2t)}&=&\frac{1+2u}{-2t}+
\frac{2\pi^2}{3}(u+1/2)(s+1/2)+\cdots\nonumber\\
&=&\frac{1}{2}+\frac{u-s}{-2t}+\frac{\pi^2}{6}(u-s-t)(s-t-u)+\cdots\nonumber\\
&=&\lim_{s,t,u\rightarrow
0}\frac{\Ga(-2t)\Ga(1-s+t+u)}{\Ga(u-t-s)}\nonumber\eeqa where in
the second line above we have used the constraint \reef{con1}. In
the third line we write the gamma functions that produce terms in
the  second line at the limit $s,t,u\rightarrow 0$. Similar thing
can be done for all other gamma functions in \reef{a717}. The
result is \beqa A^{\rm
universal}_s&\sim&\alpha\frac{\Ga(-2s)\Ga(1+s+u-t)}{\Ga(u-s-t)}+
\beta\frac{\Ga(-2s)\Ga(1+s+t-u)}{\Ga(t-s-u)}\nonumber\\
&&-
\gamma\frac{\Ga(1+s+u-t)\Ga(1+s+t-u)}{\Ga(1+2s)}\,\,,\nonumber\\
A^{\rm
universal}_u&\sim&-\alpha\frac{\Ga(1+t+u-s)\Ga(1+s+u-t)}{\Ga(1+2u)}+
\beta\frac{\Ga(-2u)\Ga(1+u+t-s)}{\Ga(t-u-s)}\nonumber\\
&&+
\gamma\frac{\Ga(-2u)\Ga(1+s+u-t)}{\Ga(s-u-t)}\,\,,\nonumber\\
A^{\rm
universal}_t&\sim&\alpha\frac{\Ga(-2t)\Ga(1+t+u-s)}{\Ga(u-t-s)}-
\beta\frac{\Ga(1+t+u-s)\Ga(1+s+t-u)}{\Ga(1+2t)}\nonumber\\
&&+\gamma
\frac{\Ga(-2t)\Ga(1+s+t-u)}{\Ga(s-t-u)}\,\,.\labell{uni1}\eeqa Now
the non-abelian  limit  is  $s,t,u\rightarrow 0$. We have changed
the superscript $A^{\rm tachyon}$  to $A^{\rm universal}$. The
reason is that the S-matrix element of massless scalars can also
be written in the above form. To see this consider the S-matrix
element of four scalar vertex operators \cite{mgjs,jp}, that
is\footnote{The S-matrix element of the  scalar vertex operators
can be read from the S-matrix element of gauge field vertex
operators by restricting the polarization of the gauge fields to
transverse directions and the momentum of the gauge fields to the
world-volume directions.} $A^{\rm scalar}=A^{\rm scalar}_s+A^{\rm
scalar}_t+A^{\rm scalar}_u$ where \beqa A^{\rm scalar}
_s&\sim&\z_1\inn\z_2\z_3\inn\z_4\left(\alpha\frac{\Ga(-2s)
\Ga(1-2t)}{\Ga(-2s-2t)}+
\beta\frac{\Ga(-2s)\Ga(1-2u)}{\Ga(-2s-2u)}\right.\nonumber\\
&&\left.-
\gamma\frac{\Ga(1-2t)\Ga(1-2u)}{\Ga(1-2t-2u)}\right)\,\,,\nonumber\\
A^{\rm scalar}
_u&\sim&\z_1\inn\z_3\z_2\inn\z_4\left(-\alpha\frac{\Ga(1-2s)
\Ga(1-2t)}{\Ga(1-2s-2t)}+
\beta\frac{\Ga(-2u)\Ga(1-2s)}{\Ga(-2u-2s)}\right.\nonumber\\
&&\left.+
\gamma\frac{\Ga(-2u)\Ga(1-2t)}{\Ga(-2u-2t)}\right)\,\,,\nonumber\\
A^{\rm scalar}
_t&\sim&\z_1\inn\z_4\z_2\inn\z_3\left(\alpha\frac{\Ga(-2t)
\Ga(1-2s)}{\Ga(-2t-2s)}-
\beta\frac{\Ga(1-2s)\Ga(1-2u)}{\Ga(1-2s-2u)}\right.\nonumber\\
&&\left.+\gamma
\frac{\Ga(-2t)\Ga(1-2u)}{\Ga(-2t-2u)}\right)\,\,,\labell{a7}\eeqa
where $\z$'s are the scalars polarization, the on-shell condition
for the scalars are $k_i^2=0$, and the Mandelstam variables
constrain to the relation \beqa s+t+u&=&0\labell{con2}\eeqa  In
this case the non-abelian limit is $s,t,u\rightarrow 0$.  Now it
is easy to check that using the constraint \reef{con2}, the
amplitude can be written in the universal form \reef{uni1} too. We
interpret the observation that  both tachyon and massless scalars
have identical amplitude, as an indication that  the on-shell
mass of tachyon does not appear in the amplitude in the universal
form. It does, however, appear in non-universal form \reef{a717}.

The non-abelian limit of the tachyon amplitude in the universal
form \reef{uni1} is $s,t,u\rightarrow 0$.   In this limit, one
finds the following massless pole and tower of contact terms:
\beqa A^{\rm universal}
&=&-4iT_p\left(\frac{(\alpha-\beta)(t-u)}{2s}+
\frac{(\beta-\gamma)(s-t)}{2u}+
\frac{(\alpha-\gamma)(s-u)}{2t}\right.\nonumber\\
&&\left.+\z(2)(\alpha+\beta+\gamma)\left(
\frac{}{}2ut+2st+2su-u^2-t^2-s^2\right)
+\cdots\right)\,\,,\labell{a0}\eeqa where we have also normalized
the amplitude by the factor of $-4T_pi$. The Zeta function is
$\z(2)=\pi^2/6$, and dots represents terms of order cubic and
more in the Mandelstam variables.   These terms are ordered in
terms of higher Zeta functions, \ie $\z(3),\z(4),\cdots$.

Now in field theory we  calculate the same S-matrix element.
Using the free action, one finds that $k_i^2=-m^2$, and the
Mandelstam variables become \beqa
s&=&-\alpha'(k_1+k_2)^2/2\,\,=\,\,-\alpha'(-2m^2+2k_1\inn k_2)/2
\,\,,\nonumber\\
t&=&-\alpha'(k_2+k_3)^2/2\,\,=\,\,-\alpha'(-2m^2+2k_2\inn k_3)/2
\,\,,\nonumber\\
u&=&-\alpha'(k_1+k_3)^2/2\,\,=\,\,-\alpha'(-2m^2+2k_1\inn
k_3)/2\,\,. \labell{fmandel} \eeqa The on-shell condition
constrains the Mandelstam variables in the relation \beqa
s+t+u&=&2\alpha'm^2\,\,.\labell{fconstraint}\eeqa Since tachyon
mass does not appear in \reef{a0}, the mass $m$ above can be
arbitrary. Using the Feynman rules for evaluating S-matrix
elements in field theory, one realizes that the massless poles in
\reef{a0} are reproduced by non-abelian kinetic term, and the
contact terms of order $\z(2)$ are reproduced by the gauge
invariant couplings in the second line below\footnote{We didn't
try gauge invariant couplings like $T^2DDTDDT$. We expect these
terms appear in the next order which includes second covariant
derivative terms.} : \beqa&& -T_p{\rm Tr}\left(\frac{}{}\frac{
(2\pi\alpha')}{2}D_aTD^aT-
\frac{(2\pi\alpha')^2}{4}F_{ab}F^{ba}\right)\,\,\nonumber\\
&&-(2\pi\alpha')^2 T_p{\rm STr}
\left(\frac{m^4}{8}T^4+\frac{m^2}{4}T^2D_aTD^aT-\frac{1}{4}
(D_aTD^aT)^2\right)\,\, \labell{a01}\eeqa  where the covariant
derivative is $D_aT=\prt_aT-i[A_a,T]$, and STr is the symmetrised
trace prescription. In reaching to this result we have used
several times the kinematic relations \reef{fmandel} and
\reef{fconstraint}. Note that the mass $m$ appears  as
coefficient of different terms. All the terms in the second line
above are of the same order of $\alpha'$.  The higher order terms
in \reef{a0} which have coefficient of higher Zeta function \ie
$\z(n)$ with $n>2$ are related to the higher covariant
derivatives of $T$ and higher power of $m$ in which we are not
interested in  the present paper.

\subsection{Two tachyons and two gauge fields amplitude}\label{scatt}

The S-matrix element of two gauge and two tachyon vertex
operators is given by \cite{mgjs,jp}  \beqa A^{\rm
tachyon}&\!\!\!\sim&\!\!\!\frac{1}{2}\z_1\inn\z_2 \left(-\alpha
\frac{\Ga(-2s)\Ga(1/2-2t)}{\Ga(-1/2-2s-2t)}-\beta
\frac{\Ga(-2s)\Ga(1/2-2u)}{\Ga(-1/2-2s-2u)}\right.\nonumber\\
&&\left.+\gamma
\frac{\Ga(1/2-2u)\Ga(1/2-2t)}{\Ga(1+2s)}\right)\nonumber\\
&&+2\alpha'\z_1\inn k_3\z_2\inn k_4\left(\alpha
\frac{\Ga(-2s)\Ga(1/2-2t)}{\Ga(1/2-2s-2t)}-\beta
\frac{\Ga(-2s)\Ga(-1/2-2u)}{\Ga(-1/2-2s-2u)}\right.\nonumber\\
&&\left.+\gamma\frac{\Ga(-1/2-2u)\Ga(1/2-2t)}{\Ga(-2t-2u)}\right)
+3\leftrightarrow 4\,\,.\labell{a718}\eeqa The Mandelstam
variables satisfy the on-shell constraint \beqa
s+t+u&=&-1/2\labell{con3}\eeqa  The non-abelian kinetic terms of
gauge field and tachyon produce Feynman amplitudes that have
massless pole in only $s$-channel,  tachyonic poles in $t$- and
$u$-channels, and some contact terms. So, in the amplitude
\reef{a718}, one should send $s\rightarrow 0$ and $t,u\rightarrow
-1/4$. This limit is consistent with the constraint \reef{con3}.
Hence, it is the non-abelian limit of the amplitude \reef{a718}.
Here again one may use the constraint \reef{con3} to change the
limit to $s,t,u\rightarrow 0$, \eg \beqa \lim_{s\rightarrow
0;t,u\rightarrow
-1/4}\frac{\Ga(-2s)\Ga(1/2-2t)}{-1/2-2t-2s}&=&\frac{1/2+2u}{-2s}+\frac{2\pi^2}{3}
(u+1/4)(t+1/4)+\cdots
\nonumber\\
&=&\frac{1}{2}+\frac{u-t}{-2s}+\frac{\pi^2}{6}(u-s-t)(t-s-u)+\cdots\nonumber\\
&=&\lim_{s,t,u\rightarrow
0}\frac{\Ga(-2s)\Ga(1-t+u+s)}{\Ga(u-t-s)}\nonumber\eeqa where we
have used the constraint \reef{con3} in the second line above.
Similarly for all other gamma functions in \reef{a718}. The
amplitude in the universal form is then \beqa A^{\rm
universal}&\!\!\!\sim&\!\!\!\frac{1}{2}\z_1\inn\z_2 \left(-\alpha
\frac{\Ga(-2s)\Ga(1-t+u+s)}{\Ga(u-s-t)}-\beta
\frac{\Ga(-2s)\Ga(1-u+t+s)}{\Ga(t-s-u)}\right.\nonumber\\
&&\left.+\gamma
\frac{\Ga(1-u+s+t)\Ga(1-t+u+s)}{\Ga(1+2s)}\right)\nonumber\\
&&+2\alpha'\z_1\inn k_3\z_2\inn k_4\left(\alpha
\frac{\Ga(-2s)\Ga(1-t+u+s)}{\Ga(1+u-s-t)}-\beta
\frac{\Ga(-2s)\Ga(-u+t+s)}{\Ga(t-s-u)}\right.\nonumber\\
&&\left.+\gamma\frac{\Ga(-u+t+s)\Ga(1-t+u+s)}{\Ga(1+2s)}\right)
+3\leftrightarrow 4\,\,.\labell{a2}\eeqa Again the reason that we
have called the above amplitude as universal amplitude is that
the S-matrix element of two gauge fields and two scalars can also
be written in this form. To see this consider this amplitude in
the standard form \cite{mgjs,jp}  \beqa A^{\rm
scalar}&\sim&\z_3\inn\z_4\left\{\frac{1}{2}\z_1\inn\z_2
\left(-\alpha \frac{\Ga(-2s)\Ga(1-2t)}{\Ga(-2s-2t)}-\beta
\frac{\Ga(-2s)\Ga(1-2u)}{\Ga(-2s-2u)}\right.\right.\nonumber\\
&&\left.\left.+\gamma
\frac{\Ga(1-2u)\Ga(1-2t)}{\Ga(1+2s)}\right)\right.\nonumber\\
&&\left.+2\alpha'\z_1\inn k_3\z_2\inn k_4\left(\alpha
\frac{\Ga(-2s)\Ga(1-2t)}{\Ga(1-2s-2t)}-\beta
\frac{\Ga(-2s)\Ga(-2u)}{\Ga(-2s-2u)}\right.\right.\nonumber\\
&&\left.\left.+\gamma\frac{\Ga(-2u)\Ga(1-2t)}{\Ga(1-2t-2u)}
\right)+3\leftrightarrow 4\right\}\,\,.\nonumber\eeqa In this case
the Mandelstam variables  satisfy the constraint  \reef{con2}. Now
it is not difficult to check that, using the constraint
\reef{con2}, the gamma functions in above amplitude can be
written in the universal form \reef{a2}.

Writing the amplitude in the universal form \reef{a2}, one finds
its non-abelian expansion  by sending the Mandelstam variables in
it to zero, \ie $s,t,u\rightarrow 0$. We are not interested in the
contact terms which have the Zeta function $\z(3)$ and more,
since they are of higher order of $\alpha'$. So in the first big
parentheses of \reef{a2} we should keep the following terms of the
gamma expansion: \beqa
\frac{\Ga(-2s)\Ga(1-t+u+s)}{\Ga(u-t-s)}&=&\frac{1}{2}+
\frac{t-u}{2s}+\z(2)(s^2-(u-t)^2)+\cdots\,\,,\nonumber\\
\frac{\Ga(-2s)\Ga(1-u+t+s)}{\Ga(t-u-s)}&=&\frac{1}{2}+
\frac{u-t}{2s}+\z(2)(s^2-(u-t)^2)+\cdots\,\,,\nonumber\\
\frac{\Ga(1-u+s+t)\Ga(1-t+u+s)}{\Ga(1+2s)}&=&1-
\z(2)(s^2-(u-t)^2)+\cdots\,\,,\nonumber\eeqa and   in the second
parentheses we should keep the following terms: \beqa
\frac{\Ga(-2s)\Ga(1-t+u+s)}{\Ga(1+u-t-s)}&=&-\frac{1}{2s}
-\z(2)(-t+u+s)+\cdots\,\,,\nonumber\\
\frac{\Ga(-2s)\Ga(-u+t+s)}{\Ga(t-u-s)}&=&-\frac{1}{2s}+\frac{1}{-u+s+t}+
\z(2)(-t+u+s)+\cdots\,\,,\nonumber\\
\frac{\Ga(-u+t+s)\Ga(1-t+u+s)}{\Ga(1+2s)}&=&\frac{1}{-u+s+t}
-\z(2)(-t+u+s)+\cdots\,\,.\nonumber\eeqa   Replacing the above
expansion for the gamma function in \reef{a2}, one finds the
following leading terms: \beqa A^{\rm
universal}&\!\!\!\!=\!\!\!\!&4i(2\pi\alpha')T_p\left\{\frac{1}{2}\z_1\inn\z_2
\left(-\frac{\alpha+\beta}{2}+\gamma\right)\right.\labell{a3}\\
&&\left.-\frac{\alpha-\beta}{2s}\left(\frac{1}{2}(t-u)\z_1\inn\z_2+2\alpha'\z_1\inn
k_3\z_2\inn k_4\right)
-\frac{\gamma-\beta}{u-t-s}\left(\frac{}{}2\alpha'\z_1\inn
k_3\z_2\inn
k_4\right)\right.\nonumber\\
&&\left.-\z(2)(\alpha+\beta+\gamma)
\left(\frac{1}{2}\z_1\inn\z_2(s^2-(u-t)^2)+2\alpha'\z_1\inn
k_3\z_2\inn k_4(-t+u+s)\right)+\cdots\right\}\nonumber\eeqa plus
terms with $3\leftrightarrow 4$. We have normalized the amplitude
at this point by the factor $4i(2\pi\alpha')T_p$.  In above
equation, dots represents contact terms that have $\z(3)$ or
higher Zeta functions. Note that the tachyon mass does not appear
in the universal amplitude \reef{a2}, nor in its above expansion.

Now in field theory,  using the fact that particle $1,\,2$ are
massless gauge fields, and $3,\,4$ are tachyon with mass $m$, the
Mandelstam variables become:  \beqa
s&=&-\alpha'(k_1+k_2)^2/2\,\,=\,\,-\alpha'(2k_1\inn k_2)/2\,\,,\nonumber\\
t&=&-\alpha'(k_2+k_3)^2/2\,\,=\,\,-\alpha'(-m^2+2k_2\inn k_3)/2\,\,,\nonumber\\
u&=&-\alpha'(k_1+k_3)^2/2\,\,=\,\,-\alpha'(-m^2+2k_1\inn
k_3)/2\,\,. \labell{fmandel2} \eeqa Conservation of momentum
constrains them in the relation \beqa
s+t+u&=&\alpha'm^2\,\,.\labell{fconstraint2}\eeqa Consider adding
the following terms to the action \reef{a01}:  \beqa -T_p{\rm
STr} \left(\pi\alpha'm^2T^2
+(2\pi\alpha')^3(-\frac{1}{8}m^2T^2F_{ab}F^{ba}+
\frac{1}{2}F_{ab}F^{bc}D_cTD^aT
-\frac{1}{8}D_aTD^aTF_{bc}F^{cb})\right)\,\,.\nonumber\eeqa It is
a simple exercise to verify that the commutator $[A_a,T][A^a,T]$
of the tachyon kinetic term reproduces the terms in the first
line of \reef{a3}. The Feynman  diagram with one vertex
$\prt_aT[A^a,T]$ from the tachyon kinetic term, the other vertex
$\prt_aA_b[A^a,A^b]$ from the kinetic term of the gauge field,
and the gauge field as propagator reproduces the massless pole in
\reef{a3}. The  diagram with the two vertices $\prt_aT[A^a,T]$
from the tachyon kinetic term and tachyon as propagator, \ie
$G_T=i/(\pi\alpha'T_p(2u-\alpha'm^2))=i/(\pi\alpha'T_p(u-s-t))$,
reproduces the tachyonic pole in \reef{a3}. All these confirm
that we expanded the amplitude correctly, \ie it produces the
Feynman amplitudes resulting from non-abelian kinetic term. It is
easy to verify  that the next leading terms in the third line of
\reef{a3} are reproduce by the contact terms above.

The  action \reef{a01} and above couplings are the leading terms
of the expansion of the following symmetrised trace non-abelian
tachyonic BI action \beqa {\cal L}^{\rm BI}&=&-T_p\,{\rm
STr}\left(\frac{}{}
V(T)\sqrt{-\det(\eta_{ab}+2\pi\alpha'F_{ab}+2\pi\alpha'
D_aTD_bT)}\right)\,\,,\labell{dbi}\eeqa  where  the tachyon
potential has the expansion \ie \beqa
V(T)&=&1+\pi\alpha'm^2T^2+\frac{1}{2}(\pi\alpha' m^2
T^2)^2+\cdots\,\,.\labell{v1} \eeqa Now it raises the question
that if one analyzes  the higher S-matrix elements, would one
find consistency  between the leading terms of the S-matrix
elements and the  above action? The above action in the abelian
case when tachyon freezes at $T=0$, reduces to the Born-Infeld
action \cite{an,esf}\footnote{It should be emphasized here that
the BI action is an action to all order of $\alpha'$. The
effective action to a fixed order of $\alpha'$, however,  has
many other terms which has less power of field strength but higher
derivative of it. Only in the situations  that the higher
derivative terms are zero, the BI action is an effective action.}
. In that case, the non-abelian extension of the action is
proposed to be the symmetrised trace of non-abelian
generalization of Born-Infeld action \cite{aat}. The leading
terms of S-matrix element of four gauge fields at the
low-energy/non-abelian limit was shown to be consistent with this
non-abelian action \cite{dg}. However, there are indication that
the symmetrised trace prescription does not work for $F^6$
\cite{ah}. Using the fact that the massless scalar couplings can
be added to the Born-Infeld action using the T-duality rules
\cite{cb}, one expects that the coupling of six scalars not to be
consistent with the symmetrised trace prescription either. Now,
extending the observation made here for four-point function that
the S-matrix element of tachyon and scalar vertex operators can
be written in the universal form, to the six-point function, one
may conclude that the symmetrised trace prescription in
\reef{dbi} does not work for the S-matrix elements of six
tachyons either. However, restricting the results to abelian case,
one may expect to have consistency with abelian tachyonic BI
action.

\section{Amplitudes in Bosonic string theory}
In this section we will analyze the  S-matrix element of four
tachyons, and the S-matrix element of two tachyons and two gauge
fields in the bosonic theory.
\subsection{Four tachyons amplitude}
The  S-matrix element of four tachyon vertex operators in the
bosonic string theory is given by \cite{mgjs,jp}  \beqa A^{\rm
tachyon} \sim-3\left(\frac{}{}\alpha B(-2s-1,-2t-1)+\beta
B(-2s-1,-2u-1)+\gamma B(-2u-1,-2t-1)\right)\nonumber\eeqa The
on-shell condition for tachyon momentum is $k^2=1/\alpha'$, and
the Mandelstam variables satisfy the constraint \beqa
s+t+u&=&-2\labell{con4}\eeqa The extra factor of -3 in the
amplitude is for later use.

To find the non-abelian limit of this amplitude, \ie the limit
that produces Feynman amplitudes resulting from non-abelian
kinetic term,  one may compare above amplitude with the S-matrix
element of four scalars. This because we know how to find the
non-abelian limit in that case. This trick was also used in
\cite{mg3} to find the limit \reef{lim1} in superstring case.

The S-matrix element of four scalar vertex operators in the
bosonic theory is given as \cite{mgjs,jp}, $A^{\rm
scalar}=A_s^{\rm scalar}+A_t^{\rm scalar}+A_u^{\rm scalar}$ where
\beqa A_s^{\rm
scalar}&\sim&\z_1\inn\z_2\z_3\inn\z_4\left(\frac{}{}\alpha
B(-2s-1,-2t+1)+\beta B(-2s-1,-2u+1)\right.\nonumber\\
&&\left.+\gamma
B(-2u+1,-2t+1)\frac{}{}\right)\nonumber\\
 A_u^{\rm scalar}&\sim&\z_1\inn\z_3\z_2\inn\z_4\left(\frac{}{}
 \alpha B(-2s+1,-2t+1)+\beta B(-2s+1,-2u-1)\right.\nonumber\\
 &&\left.+\gamma
B(-2u-1,-2t+1)\frac{}{}\right)\nonumber\\
A_t^{\rm scalar}&=& \z_1\inn\z_4\z_2\inn\z_3\left(\frac{}{}\alpha
B(-2s+1,-2t-1)+\beta B(-2s+1,-2u+1)\right.\nonumber\\
&&\left.+\gamma B(-2u+1,-2t-1)\frac{}{}\right)\labell{aa1}\eeqa
Note that $A_s^{\rm scalar}$, $A_t^{\rm scalar}$, and $A_u^{\rm
scalar}$ have tachyonic and massless poles only in $s$-, $t$-, and
$u$-channel, respectively. In this case, we know that the Feyman
amplitudes resulting from non-abelian kinetic term are reproduced
by above amplitude when expanding it at $s,t,u\rightarrow 0$.
Moreover, all the Beta functions have contribution to these
amplitudes. A minute thinking about these facts, one realizes
that to find the non-abelian limit of the tachyon amplitude, one
should write it as $A^{\rm tachyon}=A_s^{\rm tachyon}+A_t^{\rm
tachyon}+A_u^{\rm tachyon}$ where $A_s^{\rm tachyon}=A_t^{\rm
tachyon}=A_u^{\rm tachyon}$ and \beqa A_s^{\rm
tachyon}&=&-\left(\frac{}{}\alpha B(-2s-1,-2t-1)+\beta
B(-2s-1,-2u-1)+\gamma B(-2u-1,-2t-1)\right)\nonumber\eeqa The
amplitude $A_s^{\rm tachyon}$, $A_s^{\rm tachyon}$, and $A_s^{\rm
tachyon}$, should produce the massless pole in $s$-, $t$- and
$u$-channel, respectively,  that the non-abelian tachyon kinetic
 produces. The non-abelian limit is then
  \beqa s-{\rm channel}:&&\lim_{s\rightarrow
0\,,t,u\rightarrow -1}A_s^{\rm
tachyon}\nonumber\\
t-{\rm channel}:&&\lim_{t\rightarrow 0\,,s,u\rightarrow
-1}A_t^{\rm
tachyon}\nonumber\\
u-{\rm channel}:&&\lim_{u\rightarrow 0\,,s,t\rightarrow
-1}A_s^{\rm tachyon}\labell{lim2}\eeqa Note that in this limit,
for instance,  the tachyonic pole of $A_s^{\rm tachyon}$ in
$s$-channel and massless and tachyonic poles in $t$ and $u$
channels must be expanded. Similarly for $A_t^{\rm tachyon}$ and
$A_u^{\rm tachyon}$. Again the above limit can be changed to the
$s,t,u\rightarrow 0$ limit by imposing  the constraint
\reef{con4} to  rewrite the amplitude in the universal form. The
amplitude, in the universal form, in this case has  extra
tachyonic pole relative to the amplitude in the superstring case
\cite{mg2}, \ie \beqa A_s^{\rm bosonic \,
string}&=&\frac{1}{1+2s}A_s^{\rm
superstring}\,\,,\nonumber\\
A_u^{\rm bosonic \,
string}&=&\frac{1}{1+2u}A_u^{\rm superstring}\,\,,\nonumber\\
A_t^{\rm bosonic \, string}&=&\frac{1}{1+2t}A_t^{\rm
superstring}\,\,,\labell{universal}\eeqa where the $A^{\rm
superstring}$,s are given in \reef{uni1}. Using the constraint
\reef{con2}, it is easy to check that   the scalar amplitude
\reef{aa1}  can  be rewritten in the above universal form too.
Note that the amplitudes $A_s\, , A_t$, and $A_u$ are not
identical in the universal from. Writing $1/(1+2x)=-2x/(1+2x)+1$,
one may rewrite the universal amplitude as  $A^{\rm
universal}=A^{\rm superstring}+A^{\rm extra}$ where \beqa A^{\rm
extra}
_s&\sim&\frac{1}{1+2s}\left(\alpha\frac{\Ga(1-2s)\Ga(1+s+u-t)}
{\Ga(u-s-t)}+
\beta\frac{\Ga(1-2s)\Ga(1+s+t-u)}{\Ga(t-s-u)}\right.\nonumber\\
&&\left.+\gamma\frac{\Ga(1+s+t-u)\Ga(1+s+u-t)}{\Ga(2s)}\right)\,\,,\nonumber\eeqa
similar expressions  for $A_u^{\rm extra}$ and $A_t^{\rm extra}$.

The non-abelian limit of the amplitude in the universal form is
$s,t,u\rightarrow 0$. Expansion of the $A^{\rm superstring}$ part
at this limit has been already discussed in previous section. The
$A^{\rm extra}$ part has no massless pole but has tachyonic poles.
These tachyonic poles are not related to the Feynman amplitudes in
field theory that result from kinetic term. They are related to
the three tachyon couplings in field theory. However, in the
non-abelian limit, \ie $s,t,u\rightarrow 0$, these poles must be
expanded. This is what has been done  in \cite{an,esf,wt,mg4} to
find the  gauge invariant/covariant  couplings of massless fields.
Since these tachyonic poles must be expanded, they  indicate that
the non-abelian field theory may have no three tachyon coupling.
In fact, expansion of $A^{\rm extra}$ at the non-abelian limit is
\beqa A^{\rm
extra}&=&-4iT_p\left(\frac{}{}2(\alpha-\beta)(u-t)-2s(\alpha-\beta)(u-t)-4\gamma
s^2+2s^2(\alpha+\beta)\right)\labell{extra}\\
&&-4iT_p\left(\frac{}{}2(\gamma-\alpha)(s-u)-2t(\gamma-\alpha)(s-u)-4\beta
t^2+2t^2(\alpha+\gamma)\right)\nonumber\\
&&-4iT_p\left(\frac{}{}2(\beta-\gamma)(t-s)-2u(\beta-\gamma)(t-s)-4\alpha
u^2+2u^2(\beta+\gamma)\right)+\cdots\,\,,\nonumber\eeqa where dots
represents contact terms which include Mandelstam variables in
cubic form or more. Note that the above leading terms are zero in
the abelian case.  Now consider adding the following non-abelian
gauge invariant couplings to the action \reef{dbi}: \beqa {\cal
L}_1^{\rm extra} &=&(2\alpha')^2T_p{\rm Tr} \left\{\pi i
F^{ab}D_aTD_bT+ m^2TD_aTTD^aT-m^2TD_aTD^aTT\nonumber
\right.\\
&&\left.-
\frac{1}{2}D_aTD^aTD_bTD^bT+\frac{1}{2}D_aTD_bTD^aTD^bT\right\}\,\,.
\labell{dbi1}\eeqa The tachyon kinetic term   and the first term
above give the vertex function for two external tachyons and one
internal gauge field. The propagator for internal gauge field can
also be read from the gauge field kinetic term. They are \beqa
V^a_{ij}(T_1T_2)&=&(2\pi\alpha')iT_p\left(\frac{}{}(\l_1\l_2)_{ij}-
(\l_2\l_1)_{ij}\right)(1-2s)(k_1^a-k_2^a)\,\,,\nonumber\\
(G_A)^{ab}_{ij,kl}&=&\frac{\delta^{ab}\delta_{jk}\delta_{il}}
{(2\pi\alpha')^2T_p}\frac{i}{s}\,\,.\labell{vertex}\eeqa Now the
$s$-channel Feynman diagram $V(T_1T_2)G_AV(T_3T_4)$  produces the
massless pole in the first line of \reef{a0} and the first two
terms in first line of \reef{extra}. Similarly for the $t$-channel
and $u$-channel.  The other contact terms in \reef{extra} are
reproduced by the four tachyons couplings in \reef{dbi1}. It is
important to note that the above action does not have three
tachyon coupling. The fact that the non-abelian action in the
bosonic case has the couplings \reef{dbi1}, indicates that the
symmetrised trace prescription does not work in the bosonic case
even for four point function.

\subsection{Two tachyons and two gauge fields amplitude}

The S-matrix element of two tachyons and two gauge fields in the
bosonic theory is given by \cite{mgjs,jp} \beqa A^{\rm
tachyon}&\sim&(-\frac{1}{2}\z_1\inn\z_2+2\alpha'\z_1\inn
k_3\z_2\inn k_3)\nonumber\\
&&\times\left(\frac{}{} \alpha
B(-1-2s,-2t)+\beta B(-1-2s,-2u)+\gamma B(-2u,-2t)\right)\nonumber\\
&&+2\alpha'\z_1\inn k_3\z_2\inn k_4\left(\frac{}{}\alpha
B(-1-2s,1-2t)+\beta B(-1-2s,-1-2u)\right.\nonumber\\
&&\left.-\gamma B(-1-2u,1-2t)\frac{}{}\right)+3\leftrightarrow
4\,\,.\labell{a4}\eeqa The Mandelstam variables satisfy\beqa
s+t+u&=&-1\labell{con5}\eeqa As can be seen there are massless
and tachyonic poles in all $s$, $u$ and $t$ channels. A
non-trivial question is: which of these should be expanded and
which ones should be reproduced by the non-abelian gauge  theory?
To answer this question, one has to
 find the   limit that reduces the above amplitude to a series
that  has tachyonic poles, massless poles and contact terms which
are reproduced by the non-abelian gauge theory. Since the kinetic
term in the gauge theory is identical in both superstring theory
and bosonic theory, the non-abelian limit in bosonic amplitude is
similar to the non-abelian limit in the superstring case. However,
the constraint here \reef{con5} is different than the constraint
there \reef{con3}. The non-abelian limit in the present case is,
then,  $s\rightarrow 0$ and $t,u\rightarrow -1/2$. Therefore, in
this limit,  the tachyonic pole in the $s$-channel must be
expended. This is consistent with the observation made in previous
subsection  that field theory has no three tachyons coupling, at
least to the order that we consider in this paper. Similarly, in
the non-abelian limit, the massless poles in the $t$ and  $u$
channels must be expanded which indicates that field theory may
have no gauge-gauge-tachyon coupling. Now, like in the
superstring case,  one may use the constraint \reef{con5} to
change this limit to the limit $s,t,u\rightarrow 0$, that is,
\beqa A^{\rm
universal}&\sim&(-\frac{1}{2}\z_1\inn\z_2+2\alpha'\z_1\inn
k_3\z_2\inn k_3) \left(\frac{}{} \alpha
B(-1-2s,1-t+u+s)\right.\nonumber\\
&&\left.+\beta B(-1-2s,1-u+t+s)+\gamma B(1-u+t+s,1-t+u+s)\frac{}{}\right)\nonumber\\
&&+2\alpha'\z_1\inn k_3\z_2\inn k_4\left(\frac{}{}\alpha
B(-1-2s,2-t+u+s)+\beta B(-1-2s,-u+s+t)\right.\nonumber\\
&&\left.-\gamma B(-u+t+s,2-t+u+s)\frac{}{}\right)+3\leftrightarrow
4\,\,.\labell{a719}\eeqa To compare it with the S-matrix element
of two gauge and two scalar vertex operators, consider the latter
amplitude \cite{mgjs,jp},\beqa A^{\rm
scalar}&\sim&\z_3\inn\z_4\left\{(-\frac{1}{2}\z_1\inn\z_2+2\alpha'\z_1\inn
k_3\z_2\inn k_3)\right.\nonumber\\
&&\left.\times\left(\frac{}{} \alpha
B(-1-2s,1-2t)+\beta(-1-2s,1-2u)+\gamma B(1-2u,1-2t)\right)\right.\nonumber\\
&&\left.+2\alpha'\z_1\inn k_3\z_2\inn k_4\left(\frac{}{}\alpha
B(-1-2s,2-2t)+\beta B(-1-2s,-2u)\right.\right.\nonumber\\
&&\left.\left.-\gamma B(-2u,2-2t)\frac{}{}\right)+3\leftrightarrow
4\right\}\,\,,\nonumber\eeqa where the Mandelstam variables
satisfy $s+t+u=0$. Using this constraint, one can easily check
that the above amplitude can be rewritten as \reef{a719}. That is
why we call the amplitude \reef{a719} the universal amplitude.

In the universal form,  the non-abelian limit is
$s,t,u\rightarrow 0$. It is easy now to see from the  universal
amplitude \reef{a719} that the $s$-channel tachyonic pole of
amplitude in the non-universal form \reef{a4} has to be expanded
whereas the tachyonic poles in the $u$ and $t$ channels should be
reproduced by the non-abelian gauge theory.

To study the non-abelian limit of the amplitude, like in the
previous subsection, it is convenient to separate the amplitude
into two parts. One is the amplitude that has the same dependency
on the momenta as in the superstring case. We call it $A^{\rm
superstring}$. The non-abelian limit of this part is consistent
with the non-abelian  BI action \reef{dbi}. The other part  has
all other terms that appear only in the bosonic theory. We call
it $A^{\rm extra}$. Hence, \beqa A^{\rm universal}&=& A^{\rm
superstring}+A^{\rm extra}\,\,,\nonumber\eeqa where $A^{\rm
extra}$ has the following terms:\beqa && 2\alpha'\z_1\inn
k_3\z_2\inn
k_3\left(\alpha\frac{\Ga(-2s)\Ga(1-t+u+s)}{\Ga(u-t-s)}\right.\nonumber\\
&&\left.+\beta\frac{\Ga(-2s)\Ga(1-u+t+s)}{\Ga(t-u-s)}
-\gamma\frac{\Ga(1-u+t+s)\Ga(1-t+s+u)}{\Ga(1+2s)}\right)\nonumber\\
&&+2\alpha'\z_1\inn k_3\z_2\inn
k_4\left(\alpha(-t+u+s)\frac{\Ga(-2s)\Ga(1-t+u+s)}{\Ga(1+u-t-s)}\right.\nonumber\\
&&\left.+\beta(t-u-s)\frac{\Ga(-2s)\Ga(-u+s+t)}{\Ga(t-u-s)}+
\gamma(-t+u+s)\frac{\Ga(-u+t+s)\Ga(1-t+u+s)}{\Ga(1+2s)}\right)\nonumber\\
&&+\frac{1}{1+2s}\left\{(-\frac{1}{2}\z_1\inn\z_2+2\alpha'\z_1\inn
k_3\z_2\inn
k_3)\left(\alpha\frac{\Ga(1-2s)\Ga(1-t+u+s)}{\Ga(u-t-s)}\right.\right.\nonumber\\
&&\left.\left.+\beta\frac{\Ga(1-2s)\Ga(1-u+t+s)}{\Ga(t-u-s)}
+\gamma\frac{\Ga(1-u+t+s)\Ga(1-t+s+u)}{\Ga(2s)}\right)\right.\nonumber\\
&&\left.+2\alpha'\z_1\inn k_3\z_2\inn
k_4\left(\alpha\frac{\Ga(1-2s)\Ga(2-t+u+s)}{\Ga(1+u-t-s)}
+\beta\frac{\Ga(1-2s)\Ga(-u+s+t)}{\Ga(-1+t-u-s)}\right.\right.\nonumber\\
&&\left.\left.-\gamma\frac{\Ga(-u+t+s)\Ga(2-t+u+s)}{\Ga(2s)}\right)\right\}
+3\leftrightarrow 4\,\,.\labell{a6}\eeqa Expanding the gamma
functions and the tachyonic pole $1/(1+2s)$ at $s,t,u\rightarrow
0$, after some simple algebra,  one finds the following leading
terms: \beqa A^{\rm
extra}&=&-4i(2\pi\alpha')T_p\left\{-\frac{\alpha-\beta}{2s}
\left(\frac{}{}\alpha'\z_1\inn
k_2\z_2\inn k_1(t-u)\right)+\alpha'\z_1\inn k_3\z_2\inn
k_4(\alpha-\beta)\right.\nonumber\\
&&\left.-\alpha'\z_1\inn k_3\z_2\inn
k_3(\alpha+\beta-2\gamma)-2\alpha'\z_1\inn k_3\z_2\inn
k_4(\alpha-\gamma)\right.\nonumber\\
&&\left.+\frac{1}{2}\z_1\inn\z_2\left(\frac{}{}\alpha(u-t-s)+\beta(t-u-s)+2\gamma
s\right)\right.\nonumber\\
&&\left.+\left(\frac{}{}s\z_1\inn\z_2+
\alpha' k_1\inn\z_2k_2\inn\z_1\right)(\alpha-\beta)(t-u)\right.\nonumber\\
&&\left.+\left(\frac{}{}s\z_1\inn\z_2+
\alpha'k_1\inn\z_2k_2\inn\z_1\right)s(\alpha+\beta-2\gamma)
+3\leftrightarrow 4+\cdots\frac{}{}\right\}\,\,.\labell{a8}\eeqa
The above leading terms are zero for the abelian case, \ie when
$\alpha=\beta=\gamma=1$. In reaching to the above  result we have
used the conservation of momentum and the identities: \beqa
\frac{(-t+u+s)(1-t+u+s)}{(1+2s)(-u+t+s)}&=&\frac{t-s-u}{1+2s}+
\frac{s+u-t}{-u+t+s}\,\,,\nonumber\\
\frac{2s(-t+u+s)}{(1+2s)(-u+t+s)}&=&\frac{-2s}{1+2s}+\frac{2s}{-u+t+s}
\,\,.\nonumber\eeqa Note that the poles like $1/(-u+s+t)$ which
appear in \reef{a6} are disappeared  in the expanded amplitude
\reef{a8}. Now consider the non-abelian tachyonic BI action
\reef{dbi}, the couplings \reef{dbi1}, and the following
couplings \beqa {\cal L}_2^{\rm extra} &=&(2\alpha')^2\pi T_p{\rm
Tr} \left\{\frac{4\pi i}{3}
F^{ab}F_{ac}F_b{}^c+m^2TF_{ab}TF^{ab}-m^2TF_{ab}F^{ab}T\nonumber
\right.\\
&&\left.-D_aTD^aTF_{bc}F^{bc}+D_aTF_{bc}D^aTF^{bc}\right\}\,\,.
\labell{dbi2}\eeqa The gauge field kinetic term   and the first
term above give the following vertex function for two external
gauge fields and one internal gauge field:\beqa
V^a_{ij}(A_1A_2)&=&(2\pi\alpha')^2iT_p\left(\frac{}{}(\l_1\l_2)_{ij}-
(\l_2\l_1)_{ij}\right)\left\{\frac{}{}\z_1\inn\z_2(k_1^a-k_2^a)+
2k_2\inn\z_1\z_2^a-2k_1\inn\z_2\z_1^a\right.\nonumber\\
&&\left.-2s\z_1\inn\z_2(k_1^a-k_2^a)-
2k_1\inn\z_2k_2\inn\z_1(k_1^a-k_2^a)\frac{}{}\right\}\,\,.
\nonumber\eeqa Now the $s$-channel Feynman diagram
$V(A_1A_2)G_AV(T_3T_4)$, where $G_A$ and $V(T_3T_4)$ are given in
\reef{vertex},  produces the massless pole in the second line of
\reef{a3} and the massless pole in the first line of \reef{a8}.
It also produces some contact terms. These contact terms and the
contact terms $AATT$ resulting from the first term in \reef{dbi1}
reproduce the contact terms in the first four lines of
\reef{a8}.  The contact terms in the last line of \reef{a8} are
reproduced by the other couplings in \reef{dbi2}. On the other
hand, the first term in \reef{dbi1} can produce, in general,
vertex function with one gauge and one tachyon as external states
and one tachyon as internal state. So with two of these vertex
functions one produces a S-matrix element for two tachyons and
two gauge fields. However, imposing the on-shell condition for
the external gauge field, one finds that the result is zero. This
is again consistent with the string theory result \reef{a8} which
has no tachyonic pole. It is important to note that the action
\reef{dbi2} has no gauge-gauge-tachyon coupling.

\section{Discussion}

In this paper we have proposed  that there is unique
expansion/limit for tachyon S-matrix elements that corresponds to
non-abelian gauge symmetry.   The non-trivial question is then
how to find the non-abelian limit for the S-matrix elements. We
have speculated that for those S-matrix elements which, in their
expansion at the non-abelian limit, there are the Feynman
amplitudes resulting from the  non-abelian kinetic term, the
limit can easily be found.  The prescription for finding the
non-abelian limit of these S-matrix elements  has two steps: 1-
Write the S-matrix elements in the universal form. 2- Expand them
as the Mandelstam variables in them go to zero. We analyze in
details the S-matrix element of four tachyon vertex operators,
and  the S-matrix element of two tachyons and two gauge fields in
favor of this proposal. We then find the gauge invariant tachyon
action that is consistent with the leading terms of the S-matrix
element in this expansion. In the superstring theory, the
non-abelian gauge invariant action is consistent with the
symmetrised trace of the direct non-abelian generalization of the
tachyonic BI action \reef{dbi}. In the bosonic theory, the action
has some extra gauge invariant couplings that are zero in the
abelian case.

In principle, finding non-abelian  tachyon action from universal
S-matrix elements has less ambiguity than finding abelian tachyon
action from universal S-matrix elements. In the latter case, the
calculation has the ambiguity that one can replace
$\alpha'\prt_a\prt^aTf(T,A)$ by   $\alpha'm^2Tf(T,A)$ where $m^2$
is arbitrary mass, whereas, in the non-abelian case the same term
appears as $\alpha'D_aD^aTf(T,A)$ which can not be replaced by
the tachyon mass because  it represents coupling to gauge field
as well. Consistency with other S-matrix element may then fix this
arbitrariness. For example, one may change the coefficient of
$m^4T^4$ in the tachyon potential \reef{a01}, in expense of adding
coupling $m^2T^3D_aD^aT$ to this action. This ambiguity, however,
can be fixed by analyzing the S-matrix element of four tachyons
and one gauge field vertex operator. Moreover, the ambiguity is
only between the couplings that have $m$ as their coefficient.
Because those  tachyon couplings  that are independent of $m$ can
be read also from  the scalar couplings. The latter couplings have
no ambiguity. For instance, there is no ambiguity between $m^4T^4$
and $T^2D_aD^aTD_bD^bT$. Mass of tachyon does not appear as the
coefficient of this term. So if this term appears for tachyon, it
should also appear in the massless case. However, using the fact
that the massless scalars can be added into the action by
imposing the T-duality rules \cite{cb}, such term is not
consistent with the gauge symmetry. Hence, there is no such
coupling for tachyon either.

In the bosonic theory, structure of terms in \reef{dbi1} and
\reef{dbi2} are like the terms one finds in analyzing the S-matrix
element of four gauge fields \cite{an} \beqa
\Tr((4i\alpha'/3)F_{ab}[F^a{}_c,F^{bc}]+2\alpha'F^{ab}F^{cd}
[F_{ab},F_{cd}])\labell{aaaa}\eeqa
 In this present case, because
of the tachyon mass, there are some other commutators, that is,
\beqa \cL_1^{\rm extra}+\cL_2^{\rm
extra}&\!\!=\!\!&(2\alpha')^2T_p\Tr\left\{\frac{\pi
i}{2}F^{ab}[D_aT,D_bT]+\frac{2\pi^2i}{3}F^{ab}[F_{ac},F_b{}^c]+
m^2TD_aT[T,D^aT]+
\right.\nonumber\\
&&\left.+\pi
m^2TF_{ab}[T,F^{ab}]+\frac{1}{2}D^bTD_aT[D_bT,D^aT]+\pi
F^{bc}D_aT[F_{bc},D^aT]\right\}\nonumber\eeqa Since the universal
S-matrix element describes the scattering amplitude of four
tachyons or four massless scalars,  the above coupling is valid
for massless scalars too. In this case  the commutators that have
coefficient $m^2$ are zero, and the the rest can  easily  be
derived from  \reef{aaaa} using T-duality.

The leading contact terms of the off-shell S-matrix element of
four tachyons \reef{a0} is of order $(\alpha')^2$ which are
reproduced by the tachyonic BI action \reef{dbi}. The non-leading
terms of \reef{a0}  that have the Zeta function $\z(n)$ with
$n>2$ are of order $(\alpha')^n$. They  are expected to be
related to the higher derivative terms in field theory. In fact
the terms that have coefficient $\z(3)$ in the expansion of
\reef{a0} are the following: \beqa
A(\alpha'^3)&\sim&4\z(3)\left(\frac{}{}\alpha(s^2u+ut^2-su^2-u^2t)
\right.\nonumber\\
&&\left.+
\beta(u^2t+s^2t-ut^2-st^2)+\gamma(st^2+su^2-s^2t-s^2u)\frac{}{}
\right)\,\,.\nonumber\eeqa These contact  terms are zero in the
abelian case, \ie when $\alpha=\beta=\gamma=1$. So these terms
should be reproduced by some combination of non-abelian gauge
invariant coupling like $DTDDTDDTDT$ which vanishes in the
abelian case. This means in field theory at order $(\alpha')^3$
there should be no term without covariant derivative, \ie
$\z(3)(\alpha' m^2)^3T^4$. All contact terms at this order must
be reproduced by higher covariant derivative terms. The
$(\alpha')^4$ contact terms of amplitude \reef{a0} are non-zero
in the abelian case. So one can not immediately conclude that
there is no term without covariant derivative. However,  the
coefficient of these terms is $\z(4)$, so it is very unlikely
that there is  a term like $\z(4)(\alpha'm^2)^4T^4$ in field
theory. If there is such a contact term, then it  should be added
to the coefficient of $T^4$ in the expansion \reef{v1}. In that
case, the tachyon potential could not be written in  a closed
form. On the other hand, if the higher order terms in the
expansion of the string theory amplitude are reproduced only by
higher covariant derivative of the tachyon field, then \reef{v1}
is the correct expansion for the tachyon potential.  In this
case, using the fact that the on-shell tachyon potential should
vanish at the minimum of the potential, one may expect the
following form for the tachyon potential\footnote{For the
superstring theory, the on-shell tachyon potential \reef{v2} is
the same as the potential one finds in the partition function
approach \cite{aatt}.} : \beqa V(T)&=&e^{\pi\alpha'
m^2T^2}\,\,.\labell{v2} \eeqa Restricting to the abelian case, if
the mass $m$ appears only in the above potential,  then the
abelian tachyonic BI action with the above potential would be the
effective action for describing the  unstable D-branes  when
tachyon is slowly varying field. It would be interesting then to
analyze the non-leading terms of the S-matrix elements to see if
the mass $m $ does not appear in them. Structure of the higher
derivative tachyon couplings that have no $m$ as their
coefficient, are expected to be  found  also by applying
T-duality rules on the higher derivative correction to the
Born-Infeld action \cite{oda}.

%identical form for the tachyon potential in both bosonic and
%superstring theories suggests that the other massive scalars in
%the theory may have the same form for their potential. Of course
%at the minimum of the potentials they all reduce to one, whereas,
%for the tachyon it reduces to zero.

We have interpreted that the tachyon S-matrix elements in the
universal from have no reference to the mass of the tachyon vertex
operator. It indicates that there is nothing special about
tachyon vertex operators. Hence, one may expect that the S-matrix
element of any massive scalar vertex operator, in the bosonic
theory, in the class, \beqa V&=&\l\int dx (\z_i\,\prt^n
X^i)e^{ik\cdot X}\,\,,\nonumber \eeqa  to be written in the
universal form. In above equation, $n\geq 0$, $\z_i$ is
polarization of the scalar states, and the on-shell condition for
the momentum is $k^2=-(n-1)/\alpha'$. For example, the S-matrix
element of four of these vertex operators is $A=A_s+A_u+A_t$
where \beqa
A_s&\sim&\z_1\inn\z_2\z_3\inn\z_4\left(\alpha\frac{\Ga(-1-2s)
\Ga(2n-1-2t)}{\Ga(2n-2-2s-2t)}+
\beta\frac{\Ga(-1-2s)\Ga(2n-1-2u)}{\Ga(2n-2-2s-2u)}\right.\nonumber\\
&&\left.+
\gamma\frac{\Ga(2n-1-2t)\Ga(2n-1-2u)}{\Ga(4n-2-2t-2u)}\right)\,\,,\nonumber\\
A_u&\sim&\z_1\inn\z_3\z_2\inn\z_4\left(\alpha\frac{\Ga(2n-1-2s)
\Ga(2n-1-2t)}{\Ga(4n-2-2s-2t)}+
\beta\frac{\Ga(-1-2u)\Ga(2n-1-2s)}{\Ga(2n-2-2u-2s)}\right.\nonumber\\
&&\left.+
\gamma\frac{\Ga(-1-2u)\Ga(2n-1-2t)}{\Ga(2n-2-2u-2t)}\right)\,\,,\nonumber\\
A_t&\sim&\z_1\inn\z_4\z_2\inn\z_3\left(\alpha\frac{\Ga(-1-2t)
\Ga(2n-1-2s)}{\Ga(2n-2-2t-2s)}+
\beta\frac{\Ga(2n-1-2s)\Ga(2n-1-2u)}{\Ga(4n-2-2s-2u)}\right.\nonumber\\
&&\left.+\gamma
\frac{\Ga(-1-2t)\Ga(2n-1-2u)}{\Ga(2n-2-2t-2u)}\right)\,\,,\labell{aaa}\eeqa
where the Mandelstam variables are those in \reef{mandel}. They
satisfy the on-shell relation \beqa s+t+u&=&2n-2\labell{con6}\eeqa
The non-abelian limit at which the amplitude produces  massless
poles and contact terms resulting from the scalars kinetic term is
\beqa s-{\rm
channel}:&&\lim_{s\rightarrow 0\,,t,u\rightarrow n-1}A_s\nonumber\\
t-{\rm channel}:&&\lim_{t\rightarrow 0\,,s,u\rightarrow
n-1}A_t\nonumber\\
u-{\rm channel}:&&\lim_{u\rightarrow 0\,,s,t\rightarrow
n-1}A_s\nonumber\eeqa  Using the relation \reef{con6}, one can
easily rewrite the amplitude in the universal form, \ie \beqa
A_s&\sim&\frac{\z_1\inn\z_2\z_3\inn\z_4}{1+2s}\left\{
\alpha\frac{\Ga(-2s)\Ga(1+s+u-t)}{\Ga(u-s-t)}+
\beta\frac{\Ga(-2s)\Ga(1+s+t-u)}{\Ga(t-s-u)}\right.\nonumber\\
&&\left.-
\gamma\frac{\Ga(1+s+u-t)\Ga(1+s+t-u)}{\Ga(1+2s)}\right\}\,\,,\nonumber\\
A_u&\sim&\frac{\z_1\inn\z_3\z_2\inn\z_4}{1+2u}\left\{
-\alpha\frac{\Ga(1+t+u-s)\Ga(1+s+u-t)}{\Ga(1+2u)}+
\beta\frac{\Ga(-2u)\Ga(1+u+t-s)}{\Ga(t-u-s)}\right.\nonumber\\
&&\left.+
\gamma\frac{\Ga(-2u)\Ga(1+s+u-t)}{\Ga(s-u-t)}\right\}\,\,,\nonumber\\
A_t&\sim&\frac{\z_1\inn\z_4\z_2\inn\z_3}{1+2t}\left\{
\alpha\frac{\Ga(-2t)\Ga(1+t+u-s)}{\Ga(u-t-s)}-
\beta\frac{\Ga(1+t+u-s)\Ga(1+s+t-u)}{\Ga(1+2t)}\right.\nonumber\\
&&\left.+\gamma
\frac{\Ga(-2t)\Ga(1+s+t-u)}{\Ga(s-t-u)}\right\}\,\,.\nonumber\nonumber\eeqa
If one considers D$_{24}$-brane, then the scalar polarization is
$1$, and the polarization factors above can be dropped. In this
case, the above S-matrix element is exactly the amplitude for
four tachyons in the universal form \reef{universal}.  On the
other hand,  the  gauge invariant  action that we have found has
arbitrary mass. Hence, the non-abelian action is also the action
for  gauge field and the above massive scalar. When the massive
scalars are more than  one, \ie D$_p$-brane with $p<24$, the
scalar field takes the index $T^i$. It is not difficult to insert
this index in the effective action \reef{dbi}, \reef{dbi1}, and
\reef{dbi2}. The only thing that one has to do for on-shell
massive field is to replace the on-shell mass of the field into
the action. In particular the potential \reef{v2} for massive
field has no maximum, and its minimum is at zero. Whereas, the
potential for tachyon has maximum at zero and minimum at
infinity. Only in the latter case there is the interesting
physics of tachyon condensation.

One may ask the question: Is it possible to rewrite the tachyon
S-matrix element and the S-matrix element of some other massive
scalar vertex operators in the universal form?  To answer this
question, consider, for instance,  the following vertex operator:
\beqa V&=&\l\int dx\,(\z_{ij}\prt X^i\prt X^j)e^{ik\cdot
X}\,\,,\nonumber\eeqa where $k^2=-1/\alpha'$. The S-matrix
element of four of this operator has, among other things,  terms
like the following: \beqa
A&\sim&\Tr(\l_1\l_2\l_3\l_4)\Tr(\z_1^T\z_2\z_4\z_3^T)
\frac{\Ga(1-2s)\Ga(3-2t)}{\Ga(4-2s-2t)}\,\,,\nonumber\eeqa where
the Mandelstam variables are constrained in the relation
$s+t+u=2$. One can not rewrite the above S-matrix element and the
S-matrix element of four tachyons in a universal form. However,
the amplitude also has terms like \reef{aaa} in which $n=2$ and
$\z_1\inn\z_2\z_3\inn\z_4$'s are  replaced by
$\Tr(\z_1\z_2)\Tr(\z_3\z_4)$'s. These latter terms which produce
massless poles at the non-abelian limit, can be rewritten in the
universal form. Therefore, the  S-matrix element has two type of
terms. Those which can be written as universal form, have
couplings like the  tachyon couplings, and those which can not be
written in the universal form, produce other gauge invariant
couplings that the tachyon action does not have them. So one may
conclude that the non-abelian action for any arbitrary  massive
scalar state includes the tachyon action as a part it.

We have seen that for those S-matrix elements that produce at the
non-abelian limit massless and tachyonic poles that are related
to non-abelian kinetic term, the  limit can easily be found, \eg
the limit in \reef{lim2}. However, there are other S-matrix
elements that have massless and tachyonic poles that are not
related to non-abelian kinetic term. In this case, we don't know
how to find the non-abelian limit. For example, consider the
S-matrix element of three tachyons and one gauge field. This
amplitude is \beqa
A&\sim&\alpha'k_2\inn\z_1\left(\frac{}{}-\hat{\alpha}B(-1-2s,-1-2t)+
\hat{\beta}B(-1-2s,-2u)\right.\nonumber\\
&&\left.\qquad\qquad\qquad+\hat{\gamma}B(-2u,-1-2t)
\frac{}{}\right)+2\leftrightarrow 3\nonumber\eeqa where the
Mandelstam variables are given in \reef{mandel} and they satisfy
the relation $s+t+u=-3/2$. The coefficients
$\hat{\alpha},\hat{\beta}$, and $\hat{\gamma}$ are the following
group factors: \beqa \hat{\alpha}&=&\frac{1}{2}\left(\frac{}{}{\rm
Tr}(\l_1\l_2\l_3\l_4)-{\rm
Tr}(\l_1\l_4\l_3\l_2)\right)\,\,,\nonumber\\
\hat{\beta}&=&\frac{1}{2}\left(\frac{}{}{\rm
Tr}(\l_1\l_3\l_4\l_2)-{\rm
Tr}(\l_1\l_2\l_4\l_3)\right)\,\,,\nonumber\\
\hat{\gamma}&=&\frac{1}{2}\left(\frac{}{}{\rm
Tr}(\l_1\l_4\l_2\l_3)-{\rm
Tr}(\l_1\l_3\l_2\l_4)\right)\,\,.\nonumber\eeqa Note that these
group factors are zero in the abelian case. Hence, the above
S-matrix element is zero for the abelian case. The amplitude has
massless and tachyonic poles in all channels. However, neither of
them are related to only non-abelian kinetic term. The massless
poles could be reproduce by kinetic term and a gauge-gauge-tachyon
coupling, and the tachyonic poles could be reproduced by kinetic
term and a tachyon-tachyon-tachyon coupling. However, from the
study of the S-matrix element of four tachyons we have learned
that the non-abelian action can not have three tachyons coupling.
Similarly, from the study of the S-matrix element of two tachyons
and two gauge fields we have learned that the action can not have
gauge-gauge-tachyon couplings. Hence, both massless and tachyonic
poles in the above amplitude must be expanded in the non-abelian
limit, \eg $t$ must not go to 0 or -1/2. The above amplitude in
the non-abelian limit, then, produces only contact terms. These
contact terms, however, are all zero in the abelian case. In
other words, the abelian tachyon action has no
gauge-tachyon-tachyon-tachyon couplings. Since the non-abelian
tachyon kinetic term does not produce
gauge-tachyon-tachyon-tachyon amplitude, we don't know how to
find the non-abelian limit of the above amplitude.  It would be
interesting, then,  to find the non-abelian limit of these class
of S-matrix elements by other means.

{\bf Acknowledgement}: I would like to thank A. A. Tseytlin for
comments.

%\newpage

\end{document}